\documentclass[aps,twocolumn,amsmath,amssymb,floatfix]{revtex4-1}
\usepackage{graphicx}
\usepackage{dcolumn}
\usepackage{bm}
\usepackage{amsfonts}
\usepackage{graphicx}
\usepackage{bm}
\usepackage{amsmath}
\usepackage{amssymb}

\newcommand{\ty}[1]{\mbox{\tiny #1}}

\begin{document}
\title{Chirality-dependent phonon-limited resistivity in multiple layers of graphene}
\author{Hongki Min}
\author{E. H. Hwang}
\author{S. Das Sarma}
\affiliation{
Condensed Matter Theory Center, Department of Physics, University of Maryland, College Park, Maryland 20742, USA\\  
}
\date{\today}

\begin{abstract}
We develop a theory for the temperature and density dependence of phonon-limited resistivity $\rho(T)$ in bilayer and multilayer graphene, and compare with the corresponding monolayer result. For the unscreened case, we find $\rho \approx C T$ with $C \propto v_{\rm F}^{-2}$ in the high-temperature limit, and $\rho \approx A T^4$ with $A \propto v_{\rm F}^{-2} k_{\rm F}^{-3}$ in the low-temperature Bloch-Gr\"uneisen limit, where $v_{\rm F}$ and $k_{\rm F}$ are Fermi velocity and Fermi wavevector, respectively. If screening effects are taken into account, $\rho \approx C T$ in the high-temperature limit with a renormalized $C$ which is a function of the screening length, and $\rho \approx A T^6$ in the low-temperature limit with $A \propto k_{\rm F}^{-5}$ but independent of $v_{\rm F}$. These relations hold in general with $v_{\rm F}$ and a chiral factor in $C$ determined by the specific chiral band structure for a given density.
\end{abstract}

\maketitle


The hallmark, indeed the universally used definition, of a metal is its phonon-scattering induced resistivity with increasing temperature. Understanding, and if possible, controlling the electron-phonon interaction is thus one of the most important fundamental physical problems in any new electronic material. Since electron-phonon coupling typically controls the room-temperature resistivity of all metals (and doped semiconductors), the study of phonon-limited resistivity is also crucial for technological applications. In this Letter we study theoretically the electron-phonon interaction induced resistivity in graphene of \textit{arbitrary} layer thickness (including monolayer, bilayer, and multilayer graphene), finding a number of new experimentally testable results of considerable importance. 
In particular, we predict that the exponent $a$ in phonon-induced graphene resistivity, $\rho \sim T^a$,  depends on the characteristic of screening and at low enough temperatures the resistivity eventually goes as $\rho \sim T^6$ for screened phonon scattering, and that chirality has a subtle quantitative effect on the high-temperature linear-in-$T$ resistivity.

When the phonon energy ($\hbar\omega_q$) is much lower than the Fermi energy ($\varepsilon_{\rm F}$), i.e. $\hbar\omega_q \ll \varepsilon_{\rm F}$, the scattering of electrons from acoustic phonons can be divided by two regimes: $T < T_{\rm BG}$ and $T > T_{\rm BG}$. The characteristic temperature $T_{\rm BG}$ is known as the Bloch-Gr\"uneisen (BG) temperature and is defined as $k_{\rm B} T_{\rm BG} = 2 \hbar  v_{\rm ph} k_{\rm F}$, where $k_{\rm F}$ and $v_{\rm ph}$ are the Fermi wavevector and the sound velocity, respectively \cite{dassarma2010}. For $T>T_{\rm BG}$, the number of phonons increases linearly with temperature and so does the resistivity limited by electron-phonon scattering, $\rho \propto T$. This behavior in graphene has been observed experimentally \cite{chen2008,zou2010,kane1998}. In theoretical work \cite{hwang2008} it is found that the temperature dependent resistivity of monolayer graphene in BG regime is given by $\rho \propto T^4$ in the absence of screening. In a recent careful measurement of the temperature dependent resistivity of a high density graphene \cite{efetov2010}, a smooth transition of the resistivity from a linear $T$ dependence to a $T^4$ dependence is observed as the temperature decreases below $T_{\rm BG}$. The measured resistivity in Ref.~\cite{efetov2010} is the first explicit observation of BG behavior in two-dimensional (2D) systems. Graphene may be an ideal system to observe the BG behavior because at relatively high densities ($n >10^{13}$ cm$^{-2}$), $T_{\rm BG} > 200$ K and the other extrinsic scatterings are severely suppressed \cite{dassarma2010}. 
In our current work, we study a general multilayer graphene system including screening effects of the electron-phonon interaction, and demonstrate that the power-law dependence could vary smoothly from a $T^4$ to $T^6$ power law depending on the screening strength.

Motivated by the recent measurement of BG behavior of phonon limited resistivity \cite{efetov2010} we investigate the intrinsic transport properties of bilayer and multilayer graphene as limited by phonon scattering using Boltzmann transport theory in the case of $k_{\rm B} T\ll \varepsilon_{\rm F}$. i.e. high-density systems. In this paper, we consider only the longitudinal acoustic phonons because other phonon modes are negligible in the temperature range of our interest \cite{hwang2008}.

From the Boltzmann transport theory, the energy-averaged relaxation time in $k_{\rm B} T \ll \varepsilon_{\rm F}$ limit is given by \cite{ziman1960}
\begin{equation}
\label{eq:tau_average}
{1\over \left<\tau\right>}={2\pi \over \hbar} \nu (\varepsilon_{\rm F}) |C(k_{\rm F})|^2 I,
\end{equation}
where $\nu(\varepsilon_{\rm F})$ is the density of states per spin and valley at the Fermi energy $\varepsilon_{\rm F}$, $|C(k_{\rm F})|^2={D^2 \hbar k_{\rm F} \over 2\rho_{\rm m} v_{\rm ph}}$ is the squared matrix element for acoustic phonon scattering at the Fermi wavevector $k_{\rm F}$, $D$ is the acoustic phonon deformation potential, $\rho_{\rm m}$ is the graphene mass density and $v_{\rm ph}$ is the phonon velocity.
The integration factor $I$ in Eq.~(\ref{eq:tau_average}) is given by
\begin{equation}
\label{eq:integral_orig}
I=\int {d\phi\over 2\pi} {F(q)(1-\cos\phi) \over \epsilon^2(q)} {2q\over k_{\rm F}} \beta \hbar \omega_q N_q (N_q+1),
\end{equation}
where $q=2 k_{\rm F} \sin (\phi/2)$ is the magnitude of an acoustic phonon wavevector, $F(q)$ is the chiral factor defined by the square of the wavefunction projection between incoming and scattered states, $N_q=(e^{\beta \hbar \omega_q}-1)^{-1}$ is the phonon occupation number, $\omega_q=v_{\rm ph}q$ is the acoustic phonon angular frequency, $\epsilon(q)$ is a dielectric function and $\beta=1/(k_{\rm B} T)$.

Then, the conductivity in 2D system is given by
\begin{equation}
\label{eq:conductivity}
\sigma= g_{\rm s} g_{\rm v} e^2 \nu(\varepsilon_{\rm F}) {v_{\rm F}^2 \over 2} \left<\tau\right>=g_{\rm s} g_{\rm v} \left({e^2 \over h}\right)\left( {\hbar \rho_{\rm m} v_{\rm ph} v_{\rm F}^2 \over D^2 k_{\rm F} I}\right),
\end{equation}
and corresponding resistivity is $\rho=\sigma^{-1}$, where $v_{\rm F}$ is the Fermi velocity defined by $v_{\rm F}=\left. {d\varepsilon \over \hbar dk}\right|_{\varepsilon=\varepsilon_{\rm F}}$, $g_{\rm s}$ and $g_{\rm v}$ are spin and valley degeneracy factors, respectively. When several bands cross the Fermi energy, we add conductivity contributions from each crossing band (with the same screening wavevector determined from the total density of states).

The dielectric function $\epsilon(q)$ takes into account the screening effect at wavevector $q$. Within random phase approximation, $\epsilon(q)=1+q_{\rm s}(q)/q$, where $q_{\rm s}(q)$ is the screening wavelength \cite{hwang2007}. In our temperature range, we can approximate $q_{\rm s}(q)\approx q_{\rm TF}$ where $q_{\rm TF}$ is the 2D Thomas-Fermi screening wavevector defined by $q_{\rm TF}=g_{\rm s} g_{\rm v} \alpha_{\rm gr}(v/v_{\rm F}) k_{\rm F}$ where $\alpha_{\rm gr}=e^2 /(\epsilon \hbar v)$, $\epsilon$ is the effective dielectric constant and $v$ is the monolayer in-plane velocity \cite{hwang2007}.
In 2D, the strength of screening is determined by the parameter $q_0=q_{\rm TF}/k_{\rm F}\propto \alpha_{\rm gr}$ \cite{dassarma2010}, thus unscreened (strong screening) limit corresponds to $q_0, \alpha_{\rm gr}\rightarrow 0$ ($q_0, \alpha_{\rm gr} \gg 1$). 

By setting $x=q/(2 k_{\rm F})=\sin(\phi/2)$, Eq.~(\ref{eq:integral_orig}) can be reduced to
\begin{equation}
\label{eq:integral}
I={16\over \pi}\int_{0}^{1} dx { F(2 k_{\rm F} x) \over \sqrt{1-x^2}} {z_{\rm BG} x^4 e^{z_{\rm BG} x} \over (1+x_{\rm TF}/x)^2 (e^{z_{\rm BG} x}-1)^2},
\end{equation}
where $x_{\rm TF}=q_{\rm TF}/(2 k_{\rm F})$ and $z_{\rm BG}=T_{\rm BG}/T$.

In the high-temperature limit ($T\gg T_{\rm BG}$), Eq.~(\ref{eq:integral}) becomes  $I\approx z_{\infty}/z_{\rm BG}$ where 
\begin{equation}
\label{eq:integral_high_eff}
z_{\infty}={16\over \pi}\int_{0}^{1} dx {  x^4 F(2 k_{\rm F} x) \over \sqrt{1-x^2} (x+x_{\rm TF})^2 },
\end{equation}
thus the resistivity becomes $\rho \approx C T$ where 
\begin{equation}
\label{eq:gamma}
C={\pi D^2 k_{\rm B} z_{\infty} \over g_{\rm s} g_{\rm v} e^2 \hbar \rho_{\rm m} v_{\rm ph}^2 v_{\rm F}^2}.
\end{equation} 
Note that in the high-temperature limit, all phonons are thermally excited giving the linear $T$ dependence of the resistivity. Eq.~(\ref{eq:integral_high_eff}) contains the chiral factor $F$, thus $C$ depends on the chiral properties of wavefunctions. The screening effect enters only in the integration factor of $z_{\infty}$, thus does not change the temperature dependence qualitatively. Note that in the unscreened case ($\alpha_{\rm gr}=0$), $C \propto v_{\rm F}^{-2}$, while in the strong screening limit ($\alpha_{\rm gr} \gg 1$), $z_{\infty} \propto v_{F}^2$, thus $C \propto v_{\rm F}^0$, independent of $v_{\rm F}$ and density.

Next, we consider the low temperature limit ($T\ll T_{\rm BG}$). In the unscreened case, after setting $y=z_{\rm BG} x$, Eq.~(\ref{eq:integral}) becomes  
\begin{equation}
\label{eq:integral_low_unscreen}
I\approx{16 \over \pi z_{\rm BG}^4}\int_{0}^{\infty} dy {y^4 e^y \over (e^y-1)^2}={16\cdot 4! \zeta(4) \over \pi z_{\rm BG}^4}, 
\end{equation}
where $\zeta(s)$ is the Riemann-zeta function  and $F(0)=1$ was used.
Thus, the resistivity becomes $\rho \approx A T^4$ where
\begin{equation}
\label{eq:alpha_unscreen}
A={2\cdot 4! \zeta(4) D^2 k_{\rm B}^4 \over g_{\rm s} g_{\rm v} e^2 \hbar^4 \rho_{\rm m} v_{\rm ph}^5 v_{\rm F}^2 k_{\rm F}^3} \propto v_{\rm F}^{-2} k_{\rm F}^{-3}.
\end{equation}
Note that $A$ is independent of the chiral factor $F$ due to the small angle electron-phonon scattering. The $T^4$ dependence of the resistivity arises from phase space limitations for electron-phonon scattering as phonons of wavevector $2k_{\rm F}$ are no longer thermally excited. The transition from $T$ and $T^4$ dependence occurs around $T\approx T_{\rm BG}$.  

In the screened case, Eq.~(\ref{eq:integral}) becomes 
\begin{equation}
\label{eq:integral_low_screen}
I\approx
{16 \over \pi x_{\rm TF}^2 z_{\rm BG}^6}\int_{0}^{\infty} dy {y^6 e^y \over (e^y-1)^2}=
{16\cdot 6! \zeta(6) \over \pi x_{\rm TF}^2 z_{\rm BG}^6},
\end{equation}
thus the resistivity in the low-temperature limit becomes $\rho \approx A T^6$ where
\begin{equation}
\label{eq:alpha_screen}
A={2\cdot 6! \zeta(6) D^2 k_{\rm B}^6 \over (g_{\rm s} g_{\rm v})^3 e^6 \hbar^4 \rho_{\rm m} v_{\rm ph}^7 k_{\rm F}^5} \propto k_{\rm F}^{-5}.
\end{equation}
Note that compared to the unscreened case in Eq.~(\ref{eq:alpha_unscreen}), $A$ depends on $k_{\rm F}$ but not on $v_{\rm F}$, and is thus independent of a specific band structure for a given $k_{\rm F}$ or density.

\begin{figure}
\includegraphics[width=1\linewidth]{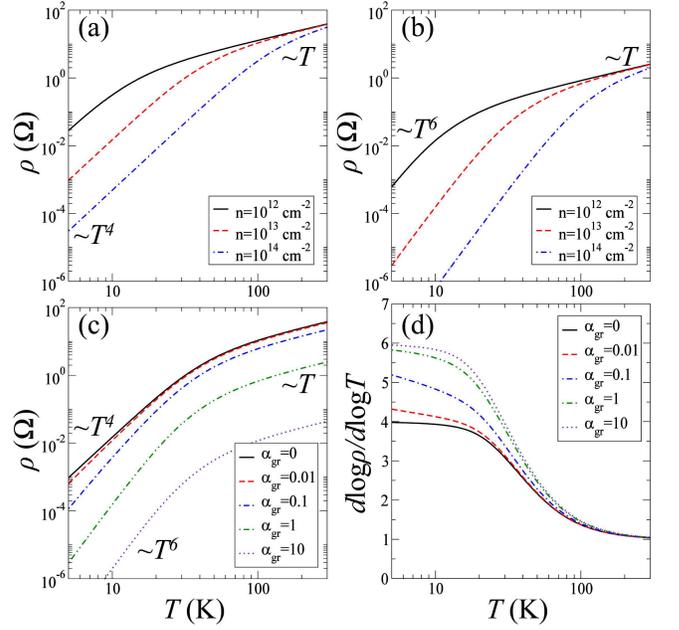}
\caption{Acoustic phonon-limited resistivity of monolayer graphene as a function of temperature for (a) $\alpha_{\rm gr}=0$ with several densities, (b) $\alpha_{\rm gr}=1$ with several densities, (c) $n=10^{13}$ cm$^{-2}$ with several $\alpha_{\rm gr}$, and (d) the logarithmic derivatives of (c).}
\label{fig:resistivity_MLG}
\end{figure}

The theory developed so far is valid for a general 2D electron system in the case of $k_{\rm B} T\ll \varepsilon_{\rm F}$, thus if we know the Fermi velocity $v_{\rm F}$ and the chiral factor $F(q)$, analytic expressions of the coefficients $A$ and $C$ can be obtained from Eqs.~(\ref{eq:gamma}), (\ref{eq:alpha_unscreen}) and (\ref{eq:alpha_screen}). Note that for monolayer graphene, $v_{\rm F}=v$ and $F(q)=(1+\cos\phi)/2$ with $q=2 k_{\rm F}\sin(\phi/2)$, whereas for bilayer graphene, 
\begin{equation}
\label{eq:velocity_bilayer}
v_{\rm F}=v {\hbar v k_{\rm F} \over \sqrt{(t_{\perp}/2)^2+(\hbar v k_{\rm F})^2}}=v\sqrt{1-\eta^2},
\end{equation}
and the chiral factor for low energy band is \cite{xiao2010}
\begin{equation}
\label{eq:chiral_factor_bilayer_l}
F(q)={1\over 4}\left[1-\eta+(1+\eta)\cos \phi\right]^2, 
\end{equation}
where $\eta=1/\sqrt{1+n/n_0}$, $n=k_{\rm F}^2/\pi$, $n_0=k_0^2/\pi$ and $\hbar v k_0=t_{\perp}/2$.

Thus, for monolayer graphene, we get
\begin{eqnarray}
\label{eq:coefficient_MLG}
C_{\ty{MLG}}(\alpha_{\rm gr}=0)&=&{\pi D^2 k_{\rm B} \over g_{\rm s} g_{\rm v} e^2 \hbar \rho_{\rm m} v_{\rm ph}^2 v^2} \propto n^0, \\
C_{\ty{MLG}}(\alpha_{\rm gr}\gg 1)&=&{2\pi \hbar D^2 k_{\rm B} \over (g_{\rm s} g_{\rm v})^3 e^6  \rho_{\rm m} v_{\rm ph}^2 } \propto n^0, \nonumber \\
A_{\ty{MLG}}(\alpha_{\rm gr}=0)&=&{2\cdot 4! \zeta(4) D^2 k_{\rm B}^4 \over g_{\rm s} g_{\rm v} e^2 \hbar^4 \rho_{\rm m} v_{\rm ph}^5 v^2 k_{\rm F}^3} \propto n^{-3/2}, \nonumber \\
A_{\ty{MLG}}(\alpha_{\rm gr}\neq 0)&=&{2\cdot 6! \zeta(6) D^2 k_{\rm B}^6 \over (g_{\rm s} g_{\rm v})^3 e^6 \hbar^4 \rho_{\rm m} v_{\rm ph}^7 k_{\rm F}^5} \propto n^{-5/2} \nonumber
\end{eqnarray}
and for bilayer graphene,
\begin{eqnarray}
\label{eq:coefficient_BLG}
C_{\ty{BLG}}(\alpha_{\rm gr}=0)&=&{1-2\eta+5\eta^2 \over 2(1-\eta^2)} C_{\ty{MLG}}(\alpha_{\rm gr}=0), \\
C_{\ty{BLG}}(\alpha_{\rm gr}\gg 1)&=&{3-10\eta+35 \eta^2 \over 8} C_{\ty{MLG}}(\alpha_{\rm gr}\gg 1), \nonumber \\
A_{\ty{BLG}}(\alpha_{\rm gr}=0)&=&{1 \over 1-\eta^2} A_{\ty{MLG}}(\alpha_{\rm gr}=0), \nonumber \\
A_{\ty{BLG}}(\alpha_{\rm gr}\neq 0)&=&A_{\ty{MLG}}(\alpha_{\rm gr}\neq 0), \nonumber \nonumber
\end{eqnarray}
where MLG and BLG stand for monolayer graphene and bilayer graphene, respectively.

Note that in the low-density limit ($n\ll n_0$ or $\eta\approx 1$), it can be shown that the 4-band bilayer results obtained here approach the 2-band low-energy bilayer results. In the high-density limit ($n\gg n_0$ or $\eta\approx 0$), 4-band bilayer $A_{\ty{BLG}}$ approaches the monolayer $A_{\ty{MLG}}$, while corresponding $C_{\ty{BLG}}$ does \emph{not} approach the monolayer $C_{\ty{MLG}}$.  

The reason for the discrepancy in $C$ between monolayer and 4-band bilayer results in the high-density limit is that in bilayer graphene, the chiral factor with the interlayer hopping $t_{\perp}=0$ cannot be obtained from the limit $t_{\perp}\rightarrow 0$ because wavefunctions involved in the two cases are different, though the energy spectra become similar. Note that the coefficient $A$ is independent of the chiral factor due to the small angle scattering, thus monolayer and bilayer results converge in high enough density.

\begin{figure}
\includegraphics[width=1\linewidth]{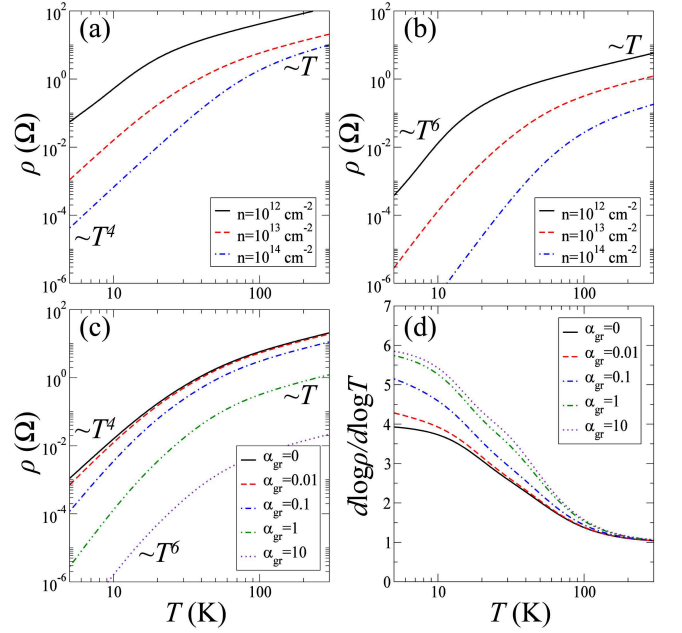}
\caption{Acoustic phonon-limited resistivity of bilayer graphene as a function of temperature for (a) $\alpha_{\rm gr}=0$ with several densities, (b) $\alpha_{\rm gr}=1$ with several densities, (c) $n=10^{13}$ cm$^{-2}$ with several $\alpha_{\rm gr}$, and (d) the logarithmic derivatives of (c).}
\label{fig:resistivity_BLG}
\end{figure}

For numerical calculations, we use $\rho_{\rm m}=7.6\times 10^{-8}$ g/cm$^2$, $v_{\rm ph}=2.6\times 10^6$ cm/s and $D=25$ eV for phonons, and the nearest-intralayer hopping $t=3$ eV and nearest-interlayer hopping $t_{\perp}=0.3$ eV. For simplicity, other remote hopping terms are neglected keeping rotational symmetry in the energy spectrum.

Figures \ref{fig:resistivity_MLG}(a) and (b) show the longitudinal acoustic phonon-limited resistivity of monolayer graphene as a function of temperature for unscreened case ($\alpha_{\rm gr}=0$) and screened case with $\alpha_{\rm gr}=1$, respectively, for several densities. For the unscreened case, the resistivity increases as $\rho \sim T^4$ at low temperatures and $\rho \sim T$ at high temperatures, while for the screened case, $\rho \sim T^6$ at low temperatures and $\rho \sim T$ at high temperatures. The transition occurs around $T_{\rm BG}$, which is given by 70.4, 222.6 and 583.6 K for $n=10^{12}, 10^{13}$ and $10^{14}$ cm$^{-2}$, respectively. Figures \ref{fig:resistivity_MLG}(c) and (d) show the temperature dependence of resistivity at $n=10^{13}$ cm$^{-2}$ for different $\alpha_{\rm gr}$ values and their logarithmic derivatives, respectively. As $\alpha_{\rm gr}$ increases from 0, the low temperature power changes from 4 and approaches to 6, as shown in Fig.~\ref{fig:resistivity_MLG}(d). Bilayer graphene results show very similar behavior, as illustrated in Fig. ~\ref{fig:resistivity_BLG}, except that the density dependence of the coefficients $A$ and $C$ are qualitatively different from the monolayer case, as expected from Eq.~(\ref{eq:coefficient_BLG}).

Let us now consider the density dependence of the coefficients $A$ and $C$ in a general multilayer graphene system.
At low energies or equivalently at low densities, arbitrarily stacked multilayer graphene is described by a superposition of pseudospin doublets \cite{min2008}, as summarized in Tab.~\ref{tab:decomposition}. Thus each pseudospin doublet contributes to the density dependence of the coefficients at low densities.
At high energies or equivalently at high densities, interlayer coupling becomes negligible and energy band structure looks like that of monolayer graphene, thus the density dependence will follow that of monolayer graphene.
\begin{table}
\caption{Chirality decomposition for periodic AB and ABC stacking up to $N=4$ layer stacks \cite{min2008}. Here we have arbitrarily labeled the first two layers starting from the bottom as A and B.}
\begin{center}
\begin{tabular}{p{0.15\linewidth} p{0.15\linewidth} | p{0.15\linewidth} p{0.15\linewidth} | p{0.15\linewidth} p{0.15\linewidth}}
\hline\hline
stacking &chirality &stacking &chirality &stacking  &chirality \\
\hline
A        &1         &ABA      &1$\oplus$2&ABAB      &2$\oplus$2\\
AB       &2         &ABC      &3         &ABCA      &4         \\   
\hline\hline 
\end{tabular}
\label{tab:decomposition}
\end{center}
\end{table}

\begin{figure}
\includegraphics[width=1\linewidth]{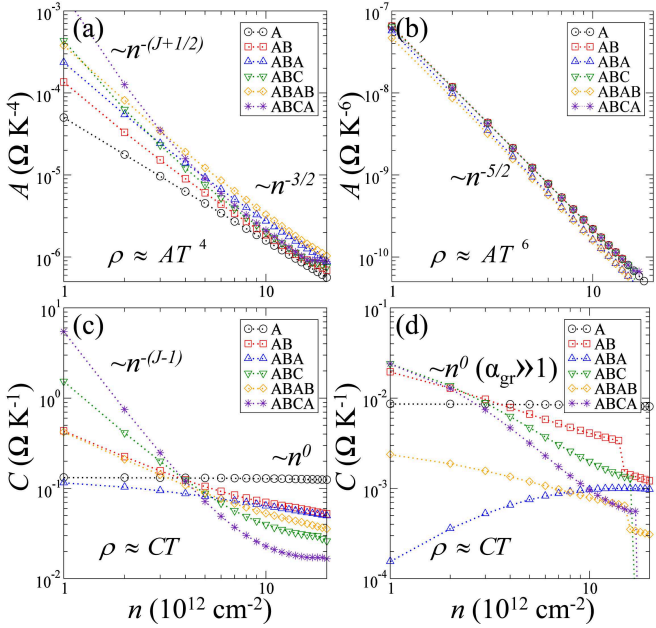}
\caption{Density dependence of (a) $A$ for $\alpha_{\rm gr}=0$, (b) $A$ for $\alpha_{\rm gr}=1$, (c) $C$ for $\alpha_{\rm gr}=0$, and (d) $C$ for $\alpha_{\rm gr}=1$ with different stacking sequences.}
\label{fig:coefficient}
\end{figure}

First, consider the density dependence of $A$ at low densities (but still assuming in $k_{\rm B}T \ll \varepsilon_{\rm F}$ limit).
Note that for $J$-chiral system, $\varepsilon_k\propto k^J$, thus $v_F\propto k_{\rm F}^{J-1}$.
From Eq.~(\ref{eq:alpha_unscreen}) and Eq.~(\ref{eq:alpha_screen}),
\begin{eqnarray}
\label{eq:alpha_dependence_J}
A(\alpha_{\rm gr}=0) &\propto& v_{\rm F}^{-2} k_{\rm F}^{-3} \propto n^{-(J+1/2)}, \\
A(\alpha_{\rm gr}\neq 0) &\propto& k_{\rm F}^{-5} \propto n^{-5/2}. \nonumber
\end{eqnarray}
Thus, at low densities, for the unscreened case ($\alpha_{\rm gr}=0$), $A$ has different density dependence depending on the chirality while for the screened case ($\alpha_{\rm gr}\neq 0$), $A$ has the same density dependence, irrespective of the chirality. 

Figure \ref{fig:coefficient}(a) and (b) show the density dependence of the coefficient $A$ for the unscreened case and screened case with $\alpha_{\rm gr}=1$, respectively, for periodic stackings up to $N=4$ layer stacks. The numerical results agree with the preceding analysis. 

Next, consider the density dependence of $C$. 
From Eq.~(\ref{eq:gamma}),
\begin{eqnarray}
\label{eq:gamma_dependence_J}
C(\alpha_{\rm gr}=0) &\propto& v_{\rm F}^{-2}\propto n^{-(J-1)}, \\
C(\alpha_{\rm gr}\gg 1) &\propto& v_{\rm F}^0\propto n^0. \nonumber
\end{eqnarray}
Thus, at low densities, for the unscreened case ($\alpha_{\rm gr}=0$), $C$ has different density dependence depending on the chirality while in the strong screening limit ($\alpha_{\rm gr}\gg 1$), $C$ is independent of density, irrespective of the chirality. Note that for monolayer graphene, $C$ remains constant with density for both unscreened and screened cases because $v_{\rm F}=v$ is constant in Eq.~(\ref{eq:gamma}). 

Figure \ref{fig:coefficient}(c) and (d) show the density dependence of the coefficient $C$ for the unscreened case and screened case with $\alpha_{\rm gr}=1$, respectively, for periodic stackings up to $N=4$ layer stacks. The numerical results agree with the preceding analysis, and at high densities density dependence of $C$ becomes weaker following the density dependence of monolayer, even though $\alpha_{\rm gr}$ is not in the strong screening limit. Note kink structures appear at a density when multiple bands begin to contribute in a multilayer stack. Also note that for ABA and ABAB stackings, multiple bands contribute to the resistivity even at low densities, thus they show relatively different behavior compared with other stackings due to the combined effects of the chirality and multiband screening. 

Even though the phonon parameters of graphene are well defined, there is uncertainty about the value of the deformation potential $D$ \cite{chen2008,efetov2010,parameter}. The results of this paper can be applied to extract the proper value of the graphene deformation potential. To get the correct value, however, it is crucial to know the significance of screening in electron-phonon scattering. Once screening is included, the larger value for $D$ is required to match the result compared with that without screening. Since two approaches give rise to the same linear-in-$T$ resistivity at high temperatures, it is hard to get the correct value of deformation potential by considering only the high-temperature result.

At low temperatures (BG regime), however, the temperature dependence of the resistivity strongly depends on screening, i.e. $\alpha_{\rm gr}$, and the inclusion of screening modifies the behavior of $\rho(T)$ from $T^4$ to $T^6$. 
In this sense, the recent experimental observation of resistivity at BG regime, $\rho \propto T^4$ around 10 K \cite{efetov2010} indicates superficially that screening in the effective electron-phonon interaction is not strong enough to show $\rho \propto T^6$ dependence. It is possible, however, that at lower temperatures the power-law behavior eventually becomes $\rho\sim T^6$, as demonstrated in Figs.~\ref{fig:resistivity_MLG}(d) or ~\ref{fig:resistivity_BLG}(d), since the screened coupling should be the appropriate interaction in an electron liquid. 
We suggest that the observed BG temperature exponent of 4 in Ref.~\onlinecite{efetov2010} is simply an effective exponent which will increase to an exponent of 6 in the asymptotic $T\ll T_{\rm BG}$ temperature regime.
For the experimental observation of 
this prediction,
it is important to work in a regime where the graphene temperature dependence from non-acoustic-phonon mechanisms \cite{non_acoustic_phonon} is quantitatively unimportant or can be subtracted out in an unambiguous manner, which may necessitate working at high carrier density and on special substrates, as has been done in Ref.~\onlinecite{efetov2010}.

Authors thank D. K. Efetov and P. Kim for helpful discussion. The work is supported by the NRI-SWAN and US-ONR.


\end{document}